# Implications of the Fourth Industrial Age on Higher Education

*Bo Xing and Tshilidzi Marwala*

**Abstract:** Higher education in the fourth industrial revolution (HE 4.0) is a complex, dialectical and exciting opportunity which can potentially transform society for the better. The fourth industrial revolution is powered by artificial intelligence and it will transform the workplace from tasks based characteristics to the human centred characteristics. Because of the convergence of man and machine, it will reduce the subject distance between humanities and social science as well as science and technology. This will necessarily require much more interdisciplinary teaching, research and innovation. This paper explores the impact of HE 4.0 on the mission of a university which is teaching, research (including innovation) and service.

## 1. The Status Quo of Our Society

Today, all graduates face a world transformed by technology, in which the Internet, cloud computing, and social media create different opportunities and challenges for formal education systems. As students consider life after graduation, universities are facing questions about their own destiny especially employment. These technologies powered by artificial intelligence are so much transforming the world that social concepts such as "post-work" are more and more defining the present period. This period requires certain skills that are not exactly the same as the skills that were required in the third industrial revolution where information technology was the key driver. These skills are critical thinking, people management, emotional intelligence, judgement, negotiation, cognitive flexibility, as well as knowledge production and management. Our starting point is to investigate the three current megatrends as well as their consequences.

### 1.1 Categories of Megatrends

We argue that one insightful lens of today's life is based on intelligent technology that is powered by artificial intelligence. Fast changes in physical (e.g., intelligent robots, autonomous drones, driverless cars, 3D printing, and smart sensors), digital (e.g., the internet of things, services, data and even people) and biological (e.g., synthetic biology, individual genetic make-up, and bio-printing) technologies, and generally in the way we work, we learn, and we live, make it a crucial force for economic competitiveness and social development.

### 1.2 The Fourth Industrial Revolution

With the waves of above mentioned breakthroughs in various domains, we gradually find ourselves in the midst of the fourth industrial revolution which is driven by artificial intelligence (AI) and cyber-physical systems (CPS) (Marwala, 2007). To understand the first industrial revolution was catalysed by Newton when he formulated his laws of motion. Because from then onwards motion was better understood and quantified, it was possible to design stem engines that mechanised much of the work that was traditionally done by humans. The second industrial revolution was catalysed by Faraday and Maxwell who unified magnetic and electric forces and this led to electricity generation and electric motor which were instrumental in the assembly lines that have come to dominate many industries. The third industrial revolution was catalysed by the discovery of a transistor which ushered the electronic age that gave us computers and internet. The fourth industrial revolution will revolutionise industries so substantially that much of the work that exists today will not exists in 50 years (Marwala et al., 2006). The next subsections describe hallmarks that characterize the fourth industrial revolution.

#### 1.2.1 Digitisation and Integration of Vertical and Horizontal Value Chains

The 4$^{th}$ industrial revolution digitises and vertically integrates processes across the entire organisation. It also integrates horizontally all the internal processes from suppliers to customers. Put simply, it epitomises a shift in paradigm shift from 'centralized' to 'decentralized' production, whereby machines no longer simple 'process' the product, but they are seamlessly integrated into the information network, the business partners and customers. In other words, the idea of consistent digitization and linking of all productive units in an economy is emphasized in the 4$^{th}$ industrial revolution age.

### *1.2.2 Digitisation of Product and Service Offerings*

Digitisation of products comprises the extension of current products, and the manufacturing of new digitised products. So far, the major gains for industrial companies have often been on improving the degree of automation but in the fourth industrial age this automation will be more intelligent and self-adaptive as more advances are made in artificial intelligence. The factory floor is moving towards self-regulating production that can be adapted to individual customer demands and has self-learning capability.

### *1.2.3 Digital Business Models and Customer Access*

Prominent industrial companies already provide disruptive digital solutions for the purpose of expanding their offerings. In the 4$^{th}$ industrial revolution era, it is possible to flexibly combine different business models with customer access (e.g., production on demand; production on site; and consumer engineering) and thereby creating new kinds of production methods. Disruptive digital business models will focus on generating extra digital revenues and optimizing customer experience in terms of interaction and access.

## 2. Introduction to Higher Education

The connection between education and society is often implied to be one-way where education is expected to fit in with economic and political trends, rather than, opposing them and representing something different. Such general understanding of the relationship between education and the socio-economic structures and what the education position involves help us to form a projection of future higher education associated with the fourth industrial revolution.

### *2.1 The History of Higher Education*

As we can observe, the development of higher education system has gone through the following stages, namely, elite, mass, and post-massification.

#### *2.1.1 Elite*

Higher education has profound origins starting in the 6$^{th}$ century monastic schools and later evolving into the medieval European University beginning in Bologna in 1088 which focused on theology and philosophy, and progressing into the current modern higher education system. In this evolution universities evolved from just being centres of teaching and learning, to include research and thereafter to include service to society. In its early stage, university education was catered for tiny elites. In that time, higher education was intended to mould the minds and characters of the ruling class.

#### *2.1.2 Mass*

In the late 20$^{th}$ century, the tension between education as a private right or a public good prompted the trend to 'massification', i.e., provide higher education to many people. During this period, the higher education spawned changes in a massive way in terms of many factors such as the size and shape of systems, the curriculum designs, the organizational structures, the pedagogy methodologies, the delivery modes, the research patterns, and the relationship between institutions and other external communities. The main goal of mass higher education was targeting transfer of skills and preparation for a wide variety of technical and economic roles.

### 2.1.3 Post-Massification

Higher education has gradually progressed from the elite phase to mass higher education and then to post-massification stages. Many advanced and some developing economies enjoy the tertiary participation rates of over 50%. Another characteristics of this trend is internationalisation of both students and staff. According to a report from OECD, with demographic changes, international student mobility is expected to reach 8 million students per year by 2025. South Africa is presently hosting many sub-Saharan students, with much of the remainder studying in European and American countries. Currently, adapting population to rapid social and technological change remains the main goal of many countries.

## 2.2 Fundamental Functions of a Higher Education Institution

The core mission of higher education remains the same whatever the era. The goal of higher education is to ensure quality of learning via teaching, to enable the students to get the latest knowledge through exploratory research, and to sustain the development of societies by means of service.

### 2.2.1 Teaching

One of the principal tasks of every university is to educate the youth. Therefore, it is necessary to implement appropriate teaching strategies and to organise work in a way that fosters learning. This has implications on adaptable learning programmes, better learning experience, and lifelong learning attitude.

### 2.2.2 Research

The journey towards global competition in the higher education requires institutions to put a huge amount of effort into research and development (R&D). Experts believe these forces range from new technology deployment to global cooperation and collaboration.

### 2.2.3 Service

To sustain the competitive position among world higher education system, we need to radically improve educational services. In particular, we need to drive much greater innovation, and competition into education.

## 3. Solutions – Higher Education in the Fourth Industrial Age (HE 4.0)

With its speed and breadth, the question brought by the aforementioned megatrends and the subsequent fourth industrial revolution for higher education it is important for nations to understand the impact of these changes to all areas of our lives including higher education.

## 3.1 Teaching in the Fourth Industrial Revolution (Teaching 4.0)

### 3.1.1 Wearables Assisted Teaching, Learning, and Training

The plurality of wearable devices produced indicates an early sign of another technology. Education establishments have to act now to realize wearables' huge potential to revolutionize the way we teach and train students and how they learn as well. Take numerical simulation, it is a very useful tool for engineers to analyse and predict the condition of real-world physical systems. In the era of the $4^{th}$ industrial revolution, when the existence of cyber-physical systems become a new norm, numerical simulations play an ever-increasing important role in both education and practical applications. Within the realm of numerical simulation, finite element analysis (FEA) is a versatile technique which has been practiced in various engineering fields such as analysing buildings (Marwala et al., 2017;

Marwala, 2012; Marwala, 2010). Modern FEA is often accomplished with the assistance of computers. As a result, students can understand key concepts more intuitively, and engineers can conduct complex modelling and interpret results easily. Nevertheless, such setup has limited the FEA processes in an entirely virtual and offline environment. These limitations in turn deprives the human perception of many physical characteristics (e.g., scale, context, spatial qualities, and materials). With the advancement of some wearable technologies, say augmented reality (AR), a user's sense and interaction with the physical world can be enhanced thereby creating a virtual laboratory. AR can supplement reality via superimposing computer-generated information over the physical context in real time which can facilitate results exploration and interpretation.

### 3.1.2  Embrace massive open online courses (MOOCs)

Teaching has long been constrained by the following scenario: students needed to gather in a lecture hall to hear the professor or sit around a table to discuss with peer fellows. Technology innovation is relaxing those constraints, however, and brining radical change to higher education. Massive open online courses, or MOOCs, is a form of education that provides stand-alone instruction online (Xing, 2015). Though much experimentation lies ahead, MOOCs threaten different universities in distinct ways. Two big factors underpin a university's costs: physical proximity requirement and productivity limitation. Because of the need for physical proximity enrolling more students is expensive considering the increase in buildings and instructors. Because of productivity limitation, the maximum number of students that can be compressed into lecture venues and exam-marking rosters are limited. MOOCs can eliminate these obstacles by working completely differently: off campus and online model; and once an online course is created, teaching extra students becomes an advantage.

### 3.1.3  Cultivating Innovative Talent

Most developing or under-developed countries lack innovative talent, especially at the high end. To fully grasp the opportunity of another wave of industrialization, a country's higher education system should not only focus on training knowledge-based skilled person, but have a good look at cultivating innovative talent, especially high-level scientists and technologists. These scientists must be trained in an interdisciplinary environment where technologists should understand humanities and social science and vice versa.

### 3.1.4  Generalize Blended Learning

Microeconomics is an important subject in higher education which has both social and practical value (Marwala, 2013; Marwala, 2014; Marwala, 2015). But most of its concepts exhibits a high level of abstraction which often imposes great difficulties for students to learn it. In many situations, the concepts are isolated, without comprehensive understanding the correlations of each piece of knowledge point on the whole picture. The aftermath of this learning process is that only parts are recognizable by students, while the comprehension of the overall working mechanism is paralysed. In this regard, the main objective for a lecturer is to let students acquire the conceptual knowledge (i.e., essential relationship between knowledge fragments and their functions in the whole knowledge system) which is applied to not only microeconomics but many other subjects as well. To address this issue, we believe a generalized blended learning (i.e., mixed e-learning and face-to-face learning methodology) may contribute to this. It is well-known that virtual environments offer great educational value in the process of information transmission and interactive participation, either in real time (e.g., video conferences), or non-simultaneous participants involvement (e.g., forums and chats). In such process, the face-to-face teaching and evaluation can be used to develop analytical expressions and problem solving capabilities related to mathematical matters. Lecturers at this stage can get physical feedback about the effectiveness of their knowledge transmission to students. Then the understanding of some specific conceptual issues are further assessed and reinforced via online graphic representations and multiple choice test questions and this offers students an advantage of reviewing their results immediately.

In closing rather than fighting against these new technologies and the associated novel teaching patterns, higher education systems need to look at how they can accept them and transform the teaching and learning environment to the benefit of both students and academics.

### *3.2 Research in the Fourth Industrial Revolution (Research 4.0)*

#### *3.2.1 Open Innovation*

Open innovation, refers to the combination of humans and computers to form distributed systems for the purpose of accomplishing innovative tasks that neither can be done alone. Despite the debate about accuracy, information science has begun to build on some early successes (say, Wikipedia) to demonstrate the potential of evolving open innovation that can model and resolve wicked problems at the junction of economic, environmental, and socio-political systems. A typical open innovation process includes:

(a) Micro-tasking under crowdsourcing mechanism where the respective strengths of a crowd and machines can be magnified.
(b) Designated workflows guide crowd-workers to use and augment the information offered by workers at the previous step.
(c) To create problem-solving ecosystems, researcher can then combine the cognitive processing of many ordinary contributors with machine-based computing to establish faithful models of the complicated, interdependent systems that underlie the world's most demanding tasks.

#### *3.2.2 Evolutionary & Revolutionary Innovations*

Under higher education in the fourth industrial age, a country's higher education system should put innovation, both evolutionary and revolutionary, high on its agenda. In general, innovations based on existing technologies are so-called evolutionary type; while revolutionary type of innovations focuses are inventions of new technologies. Ideally, hybrid innovation is a sound strategy but it is difficult to implement. Established academics are often victims of their own accomplishments. Leading scholars have long succeeded by exploring new research domains that could lead to incremental research output growth. Emerging researchers have aggressively followed a similar strategy. As one research area matures and competition increases severely, the degree of research outputs being published in the form of patents or journals inevitably gets very low. Introducing new research directions means going up against entrenched competition (Xing and Gao, 2014). In the era of the $4^{th}$ industrial revolution, higher education needs to deepen its technology system reforms by breaking down all barriers to innovation. One noteworthy obstacle is resource allocation for funding different research projects. For those technology innovations that are important for industrialization, re-industrialization, and neo-industrialization, but are unable to profit in the marketplace in the near-term, financial support from institution and government levels should be made available. However, for applied technologies where commercialization is possible, social capital can play an active role (Xing, 2017). Additionally, several other hindrance should also be dealt with properly: First, with its hybrid innovation strategy, higher education practitioners need to have a global perspective. The trend of world technology development should be well-perceived and thus appropriate plans need to be made. Each stream of innovation resources, internally, locally, regionally, and globally, should be utilized properly. Second, by having various development strategies and incentive policies across different departments, the connectivity among them should be optimized to avoid potential overlapping. Third, the speed of technology transfer need to be raised to boost the economic and social development.

#### *3.2.3 New Technological Advancement Driven Research and Development*

New technological advancements are often ranked as the most important driving force for R&D. Technology-driven R&D comes in many forms and it can mean employing mobile capabilities to improve data acquisition accuracy; using advanced big-data analytics to spot hidden statistical patterns; harnessing artificial intelligence techniques to retool information search, collection, organization, and knowledge discovery, to name just a few. The bottom line, in all cases, is that the advanced technologies can be leveraged across many domains to continue to deliver impact. Briefly, advanced technologies can bring benefits to higher education R&D in at least four areas: cost and timeline reduction; operation transformation; R&D process enhancement; and, most significant, research direction innovation via the creation of new ideas and theories. Take the example of additive manufacturing (or 3D printing), this new technology can be used to reduce the cost of producing prototypes, which are generally time consuming and cost inefficient in conventional higher education R&D. This innovation results in both significant efficiencies and more flexible experimental plans which, in turn, lead to uses of the technology where cost had previously been prohibitive in the laboratory environment. In the fourth industrial age, R&D processes with the help of advanced technologies treat functions such as IT and analytics as 'centers of value' rather than of service or cost; nurture partnership attitude; and, more and more frequently, form an agile style of R&D. One of the most important and difficult task, is to shift higher education R&D culture from an outdated 'waterfall approach' to idea development. Higher education institutions that make this change will become good at absorbing ideas from all kinds of sources.

### 3.2.4 Shorten Innovation Cycles

Speed enables higher education institutions to be aware of research trends as they emerge and catch up with competition. In comparison to commercial R&D house, higher education institutions' overly-long development times are the most-blamed obstacle to generating positive returns on innovation. In practice, fast movers are much more likely to also be strong innovators as they are also more disruptive. Brainstorming, conceptualization, model design, theoretical proving, experiment setting-up, components procurement, prototyping, test conducting, results analysis, and deliverables submission can be organized into teams that work closely with group leaders to quicken responsiveness to emerging research trends.

In closing, as we have indicated herein, the strongest innovators and leading researchers draw on swiftness, well-pruned processes, and the exploitation of advanced technology to explore and capture research opportunities. Any higher education institution thinking about research in the fourth industrial revolution should first determine where its gaps are vis-à-vis the areas mentioned above and make a plan to address those issues. Once the internal house is in an opposite situation, they can begin to scout around for attractive odds for incremental research capacity growth.

### 3.3 Service in the Fourth Industrial Revolution (Service 4.0)

#### 3.3.1 University-as-a-Platform (UaaP)

We are all aware of the downfall of Blackberry. The Red Queen effect, the requirement of running faster just to remain in the same place, is one of the most commonly cited causes. In such scenario, one rival that successfully adopts a platform-oriented methodology can compress an entire sector's innovation life cycle. The outcome of the Red Queen effect is that it becomes harder for a competitor to get ahead of the dominant player. In systems thinking, this phenomenon is often referred to escalation. Nowadays, platform concepts are creating an entirely new competition landscape, one that puts ecosystems in face-to-face wrestling. In Service 4.0, the ongoing transformation to platform-based competition is led by many forces: educational activities; ubiquitous computing and Internet of things both within and outside campus and the demanding students in terms of customized learning. That which served institutions well in scientific disciplines and speciality based markets can become their impediments in platform-based environments. Managing platform-based higher education businesses require a complete different mind-set for strategy. In non-education sector, the likes of Alibaba, eBay, Facebook, WeChat, Google, Baidu and Amazon are actively building their empires around the idea of platform thinking. While platform thinking is not new, what is new is that platform-centric styles are turning into the engines of innovation that are spreading a wide variety of unanticipated sectors such as automobiles, manufacturing, fashion, healthcare, publishing, and many others. In principle, the platform-based business models emphasize more biologically inspired thinking style rather than a mere organizing logic. In a broader view and analogic metaphor manner, higher education institutions need to reconceive their business ecosystems, re-identify their competitive edges, reshuffle their customer pools, reshape themselves as orchestrators, and rebuild service architecture. University-as-a-Platform (UaaP) gives the current higher education system an opportunity to steer their bread-and-butter businesses towards platform businesses for a better service performance. Key drivers for a successful UaaP include: a) deliver inter-, multi-, and across-disciplinary degrees; b) an appropriate blend of service models (e.g., blended learning, MOOCs, etc.); c) the emergence of Internet of everything; d) the integration of routine education activities into software across a plethora of institution system; e) up-to-date digital infrastructure; and f) enhanced connectivity among all parties residing in higher education value chain.

### 3.3.2 Education-as-a-Service (EaaS)

Typically, in the age of $4^{th}$ industrial revolution, once every couple of decades, a disruptive new technology arises that essentially changes the blueprint of many sectors. In terms of higher education, the massive proliferation of affordable mobile devices, Internet broadband connectivity and rich education content start a trend of transforming how education is delivered. Cloud computing, amongst other techniques, creates a new way of educating people that might eventually disrupts the existing higher education systems. With the support of education cloud, government decision makers and business practitioners can answer some key strategic questions comprehensively: deliver education in the quickest, most efficient and best affordability form; develop $21^{st}$ century students' skills and prepare students for the new job market in the most appropriate way; encourage native innovation with the strongest incentives; and share resources across institutions, districts, regions, or the entire country in the smoothest fashion. When universities think of embracing EaaS, they often imagine profound advertising campaigns, big promotion budgets, and huge amount of infrastructure investment. Fortunately, EaaS has a healthier respect for the students than academicians have for disruptive ways of delivering education service. At the heart of EaaS is the belief that students' needs should be met effectively. Therefore, when a higher education institution sets out to attract a potential student as a customer, it needs to create an all-round education experience that is genuinely capable of satisfying the customer's needs, although, this process is not as simple as it may seem. EaaS is not the creation of pseudo differences via a change in logo, location, or making vague promises with empty sounding words. Furthermore, higher education institutions are accountable to a host of stakeholders such as governments, accrediting agencies, the public and private funding sources, academics, management, support staff, and students. An EaaS orientation that translates into an effective education scheme will achieve these broader concerns. Nevertheless, many institutions adopt EaaS strategy poorly by giving lip service to various stakeholders. Education and technology has advanced over the past few decades. Many technology-assisted / enhanced educational practices are no longer as simplistic. In Service 4.0, EaaS as a guideline has to discover newer and more advanced strategies to cope with ever-increasing societal complexity.

### 3.3.3 Internationally-linked Programmes

With the fast pace of the $4^{th}$ industrial revolution, forging various kinds of institutional linkages, both domestically and internationally, to offer more versatile degree programmes and professional qualifications becomes a must. Among these schemes, the following types stand out and are worth consideration: First, twinning programmes where a local education provider collaborates with a foreign education provider to develop a connected system allowing course credits to be taken in different locations. On completion of the twining programme, foreign education provider awards a qualification. Second, franchise programmes is a scenario where foreign education provider authorizes a local education provider to deliver their courses / programmes, and the qualification is awarded by the foreign education provider. Third, double or joint degree is an arrangement where local and foreign education providers cooperate to offer a programme for a qualification that is awarded jointly or from each of them. Fourth, blended learning where local and foreign education providers deliver programmes to enrol students in various mixed forms, e.g., e-learning, online learning or on-site learning.

In closing we trust that improving the quality of service in higher education can bring about a significant change in the society.

### 4. Conclusions

Though the business of higher education remains unchanged since the times of Aristotle, today students still assemble at a scheduled time and venue to listen to the wisdom of scholars. Given the fourth industrial revolution, a new form of a university is emerging that does teaching, research and service in a different manner. This university is interdisciplinary, has virtual classrooms and laboratories, virtual libraries and virtual teachers. It does, however, not degrade educational experience but augment it.